\documentclass[aps,prd,showpacs,twocolumn,superscriptaddress,nofootinbib,preprintnumbers]{revtex4-1} 

\makeatletter
\def\p@subsection{}
\makeatother

\usepackage{color,graphicx,amsmath,amssymb,lipsum}
\usepackage[colorlinks=true,citecolor=blue,linkcolor=blue,urlcolor=blue, backref=false,pdfborder={0 0 0}]{hyperref}

\usepackage[normalem]{ulem}
\usepackage[utf8]{inputenc} 
\usepackage{mathtools}
\usepackage{tabularx} 
\usepackage{tabu}
\usepackage{multirow}
\usepackage[table]{xcolor}
\usepackage{colortbl}
\usepackage{caption}

\usepackage{float}

\usepackage{bm}

\usepackage[compat=1.1.0]{tikz-feynman}
\usepackage{tikz}
\usetikzlibrary{shapes.geometric}

\tikzset{
bgraviton/.style={decorate, decoration={snake, amplitude=0.5mm, segment length=2mm, pre length=.0mm, post length=.2mm}, double, thick},
source/.style={decorate, thick, blob, scale = 0.37},
hsource/.style={decorate, thick, blob, scale = 0.5},
graviton/.style={decorate, decoration={complete sines, amplitude=1mm, segment length=2mm, pre length=0mm, post length=.2mm}, thick},
myscalar/.style={decorate, dashed, thick},
binsertion/.style={decorate, thick, crossed dot},
Qcorr/.style={decorate, ultra thick, dot},
gvert/.style={
},
myheavysource/.style={
    blob,    
    scale=1, 
    on foreground layer,
    append after command={
      (\tikzlastnode) node[circle, fill=white, scale=0.5] {$H$}
    }       
  }
}

\newcommand{\scalarprop}{
\begin{tikzpicture}[baseline = {(0, -0.5)}]
        \begin{feynman}
          \vertex (top) at (0, 1);
          \vertex (i) at (-1, 0);
          \vertex (f) at (1, 0);
          \diagram*{
            (i) -- [myscalar,momentum = {[arrow shorten = 0.3] $k$}]  (f) 
          };
        \end{feynman}
    \end{tikzpicture}
}

\newcommand{\backgroundone}{
\begin{tikzpicture}
        \begin{feynman}
          \vertex (top) at (0, 1);
          \vertex (i) at (-1, 0);
          \vertex (c) at (0, 0);
          \vertex (f) at (1, 0);
          \vertex[binsertion] (c) at (0,0) {}; 
          \diagram*{
            (i) -- [myscalar] (c) -- [myscalar] (f) 
          };
        \end{feynman}
    \end{tikzpicture}
}

\newcommand{\backgroundonevertex}{
\begin{tikzpicture}[baseline = {(0,-0.4)}]
        \begin{feynman}
          \vertex (top) at (0, 1);
          \vertex (i) at (-1, 0);
          \vertex (c) at (0, 0);
          \vertex (f) at (1, 0);
          \vertex[binsertion] (c) at (0,0) {}; 
          \diagram*{
            (i) -- [myscalar,momentum = {[arrow shorten = 0.2] $k_1$}] (c) -- [myscalar,reversed momentum = {[arrow shorten = 0.2] $k_2$}] (f) 
          };
        \end{feynman}
    \end{tikzpicture}
}

\newcommand{\backgroundtwo}{
\begin{tikzpicture}
        \begin{feynman}
          \vertex (top) at (0, 1);
          \vertex (i) at (-1, 0);
          \vertex (c1) at (-0.333, 0);
          \vertex (c2) at (0.333, 0);
          \vertex (f) at (1, 0);
          \vertex[binsertion] (c1) at (-0.333,0) {}; 
          \vertex[binsertion] (c2) at (0.333,0) {}; 
          \diagram*{
            (i) -- [myscalar] (c1) -- [myscalar] (c2) -- [myscalar] (f) 
          };
        \end{feynman}
    \end{tikzpicture}
}

\newcommand{\backgroundQQ}{
\begin{tikzpicture}
        \begin{feynman}
          \vertex (top) at (0, 1);
          \vertex (i) at (-1, 0);
          \vertex (c1) at (-0.333, 0);
          \vertex (c2) at (0.333, 0);
          \vertex (f) at (1, 0);
          \vertex[Qcorr] (c1) at (-0.333,0) {}; 
          \vertex[Qcorr] (c2) at (0.333,0) {};
          \diagram*{
            (i) -- [myscalar] (c1) -- [myscalar, double ,double ] (c2) -- [myscalar] (f) 
          };
        \end{feynman}
    \end{tikzpicture}
}

\newcommand{\backgroundQQvertex}{
\begin{tikzpicture}[baseline = {(0,-0.5)}]
        \begin{feynman}
          \vertex (top) at (0, 1);
          \vertex (i) at (-1, 0);
          \vertex (c1) at (-0.333, 0);
          \vertex (c2) at (0.333, 0);
          \vertex (f) at (1, 0);
          \vertex[Qcorr] (c1) at (-0.333,0) {}; 
          \vertex[Qcorr] (c2) at (0.333,0) {};
          \diagram*{
            (i) -- [myscalar, momentum = {[arrow shorten = 0.15] $k_1$}] (c1) -- [myscalar, double ,double] (c2) -- [myscalar, reversed momentum = {[arrow shorten = 0.15] $k_2$}] (f) 
          };
        \end{feynman}
    \end{tikzpicture}
}

\newcommand{\backgroundQQoneR}{
\begin{tikzpicture}
        \begin{feynman}
          \vertex (i) at (-1, 0);
          \vertex (c1) at (-0.333, 0);
          \vertex (c2) at (0.333, 0);
          \vertex (f) at (1.7, 0);
          \vertex[Qcorr] (c1) at (-0.333,0) {}; 
          \vertex[Qcorr] (c2) at (0.333,0) {};
          \vertex[binsertion] (c3) at (1.0,0){};
          \diagram*{
            (i) -- [myscalar] (c1) -- [myscalar, double ,double ] (c2) -- [myscalar] (c3) -- [myscalar] (f) 
          };
        \end{feynman}
    \end{tikzpicture}
}

\newcommand{\backgroundQQoneL}{
\begin{tikzpicture}
        \begin{feynman}
          \vertex (i) at (-1.7, 0);
          \vertex (c1) at (-0.333, 0);
          \vertex (c2) at (0.333, 0);
          \vertex (f) at (1.0, 0);
          \vertex[Qcorr] (c1) at (-0.333,0) {}; 
          \vertex[Qcorr] (c2) at (0.333,0) {};
          \vertex[binsertion] (c3) at (-1.0,0){};
          \diagram*{
            (i) -- [myscalar] (c3) -- [myscalar] (c1) -- [myscalar, double ,double ] (c2) -- [myscalar] (f) 
          };
        \end{feynman}
    \end{tikzpicture}
}

\usepackage{ulem}
\usepackage{diagbox}

\usepackage{mathrsfs}

\newcommand{\be}{\begin{equation}}
\newcommand{\ee}{\end{equation}}
\newcommand{\beqa}{\begin{eqnarray}}
\newcommand{\eeqa}{\end{eqnarray}}

\newcommand\g{\gamma}

\renewcommand\l{\lambda}

\def\d{\partial}
\newcommand{\bseq}{\begin{subequations}}
\newcommand{\eseq}{\end{subequations}}

\renewcommand{\ln}{\mathop{\rm ln}\nolimits}

\definecolor{cornellRed}{HTML}{B31B1B}
\definecolor{cornellBlue}{HTML}{0068AC}
\definecolor{cornellGreen}{HTML}{6EB43F}
\definecolor{purple}{HTML}{66023C}

\def\gsim{\raise0.3ex\hbox{$\;>$\kern-0.75em\raise-1.1ex\hbox{$\sim\;$}}}
\def\lsim{\raise0.3ex\hbox{$\;<$\kern-0.75em\raise-1.1ex\hbox{$\sim\;$}}}

\def\beqn#1{\begin{equation}\label{#1}}
\def\eeqn{\end{equation}}

\def\beqa#1{\begin{eqnarray}\label{#1}}
\def\eeqa{\end{eqnarray}}

\def\Z2{$\mathcal{Z_2}$}

\newcommand{\Cs}[1]{{C_{#1}}}

\newcommand {\ignore}[1]{}

\begin{document}

\preprint{MIT-CTP/5664}

\title{
Gravitational Raman Scattering 
in Effective Field Theory: \\
a Scalar Tidal Matching at $\mathcal{O}(G^3)$
}

\author{Mikhail M. Ivanov}
\email{ivanov99@mit.edu}
\affiliation{Center for Theoretical Physics, Massachusetts Institute of Technology, 
Cambridge, MA 02139, USA}
\author{Yue-Zhou Li}
\email{liyuezhou@princeton.edu}
\affiliation{Department of Physics, Princeton University, Princeton, NJ 08540, USA}
\author{Julio Parra-Martinez}
\email{jpm@phas.ubc.ca}
\affiliation{Department of Physics and Astronomy, University of British Columbia, Vancouver, V6T 1Z1, Canada}
\author{Zihan Zhou}
\email{zihanz@princeton.edu}
\affiliation{Department of Physics, Princeton University, Princeton, NJ 08540, USA}

\begin{abstract} 
We present a framework to compute amplitudes 
for the gravitational analog 
of the Raman process, a
quasi-elastic scattering of waves
off compact objects, in worldline effective field theory (EFT). 
As an example, we calculate third post-Minkowskian (PM) order ($\mathcal{O}(G^3)$), or two-loop, phase shifts
for the scattering of a massless scalar field including all tidal effects and dissipation. 
Our calculation unveils 
two sources of the 
classical renormalization-group flow
of dynamical Love numbers: a universal running independent of the nature of the compact object, and a running self-induced by tides.  
Restricting to the black hole case,
we find that our EFT phase shifts 
agree exactly with those from
general relativity, provided that the relevant static Love numbers
are set to zero. 
In addition, 
we carry out a complete matching
of the leading scalar dynamical 
Love number
required to renormalize 
a universal 
short scale divergence in the 
S-wave.
Our results pave the way for systematic 
calculations of gravitational Raman 
scattering 
at higher PM orders.
\end{abstract}

\maketitle

\textit{Introduction.}--
Recent advances in gravitational wave astronomy 
have spurred the development of efficient techniques for
precision calculations of binary dynamics. One such technique 
is worldline effective field theory (EFT) for compact binaries~\cite{Goldberger:2004jt,Goldberger:2005cd,Goldberger:2007hy,Porto:2016pyg,Goldberger:2022ebt,Goldberger:2022rqf}, wherein a compact object (a neutron star or black hole) 
is represented at large distances
as a point particle, and
which provides a systematic 
program for the perturbative computation of inspiral waveforms. More generally, the EFT paradigm enables an accurate description of a variety of physical effects:
tides and dissipation~\cite{Goldberger:2005cd,Goldberger:2020fot, Goldberger:2020wbx},
spin~\cite{Porto:2007qi,Porto:2016pyg,Levi:2018nxp},
Hawking radiation~\cite{Goldberger:2019sya,Goldberger:2020geb}, 
self-force~\cite{Galley:2008ih,Galley:2010xn,Zimmerman:2015rga,Cheung:2023lnj}, etc. 

In this \textit{Letter}, 
we use the EFT framework to calculate mildly
inelastic gravitational scattering of massless 
fields off compact objects.
This is a direct gravitational 
analog of Raman scattering of photons that is 
commonly used to elucidate 
the internal structure of molecules.
Here we explore its gravitational 
counterpart to probe the nature 
of compact relativistic objects.

In the worldline EFT the finite-size structure of compact objects
is captured  by multipole moments on the particle's worldline non-minimally coupled to the gravitational field \cite{Goldberger:2004jt,Goldberger:2005cd}. The associated Wilson coefficients provide a gauge-invariant definition of the tidal deformability of the objects, also known as Love numbers~\cite{Damour:1982wm,Damour:2009vw, Damour:1998jk, Damour:2009va, Binnington:2009bb,  Goldberger:2004jt,Kol:2011vg,Hui:2020xxx,Charalambous:2021mea}. These are free parameters in the EFT which have to be either measured from data or 
extracted from a matching calculation to a microscopic theory, if the latter
is available. Once the values of matching coefficients are determined they can be used to make further predictions.  The universality and consistency of the EFT thus guarantee its
predictability.

Scattering amplitudes are particularly suitable for matching calculations:
they are simple, manifestly 
gauge-invariant, and field-redefinition independent objects~\cite{Goldberger:2004jt,Goldberger:2007hy, Bautista:2021wfy,Saketh:2022wap,Bautista:2022wjf,Ivanov:2022qqt,Saketh:2023bul}. 
In addition, in the post-Minkowskian (PM) regime (formal 
perturbation theory in Newton's constant $G$)
they can be directly compared to 
known amplitudes 
in full classical general relativity (GR). 
These matching calculations also provide new insights into the general structure 
of gravitational scattering amplitudes by confronting them with exact non-perturbative 
results from black hole solutions. 
In this vein, partial results on the calibration of Love numbers 
from scattering amplitudes 
exploiting the so-called near-far 
factorization were given in~\cite{Ivanov:2022qqt,Saketh:2023bul}.
A numerical estimation of tidal effects from scattering 
of a pointlike particle with scalar charge by black holes at 4PM order was carried out  in~\cite{Barack:2023oqp}.
Finally, the scattering of photons and gravitons 
off compact objects is, in principle, an observable phenomenon
relevant in astrophysics and cosmology, see e.g.~\cite{Nambu:2015aea,Turyshev:2017pgm,Nambu:2019sqn,Howard:2023krm}.

We present a general framework for systematic computations of 
EFT amplitudes for gravitational 
Raman scattering at high PM orders.
Our approach makes use of 
the background field method and advanced multiloop integration techniques.
We demonstrate its power by 
explicitly calculating the amplitudes
for spin-$0$ fields scattering 
off a non-spinning compact object through 
3PM order, $\mathcal{O}(G^3)$, where finite-size effects first appear. 
We find that the amplitude exhibits ultraviolet (UV) divergences, whose renormalization requires contact worldline operators. They are scalar analogs of the  ``dynamical Love number,'' 
a coefficient that sets the strength of the 
multipole moment tidally induced by an external 
time-dependent field. 
We show that dynamical Love numbers 
undergo
renormalization group 
running due to two different effects. 
The first source of renormalization
is the gravitational ``dressing'' 
of the point particle action. As such, 
this running is universal for any 
compact object. 
The second source of the running is
the gravitational ``dressing'' 
of the static Love number.
We call such running ``self-induced'',
as its strength is set by the  
amplitude of lower order 
tidal Wilson coefficients (see also~\cite{Blanchet:1997jj,Saketh:2023bul,Mandal:2023hqa,Jakobsen:2023pvx} for similar discussions).

Assuming that 
a compact object is a black hole, 
and using results from black hole perturbation theory (BHPT)~\cite{1978ApJS...36..451M,1988sfbh.book.....F,Mano:1996vt,Mano:1996mf,Mano:1996gn,Sasaki:2003xr,Dolan:2008kf,Bonelli:2021uvf,Ivanov:2022qqt,Bautista:2023sdf}, the EFT scattering amplitudes allow for a complete order-by-order matching of tidal effects, including dissipation. 
Matching the 3PM scattering  amplitudes to BHPT, we prove explicitly that the leading static tidal coefficient is zero and does not run, in agreement with previous off-shell calculations~\cite{Kol:2011vg,Hui:2020xxx,Charalambous:2021kcz}.
This also implies the 
vanishing of the self-induced 
tidal coefficients.
In addition, we completely 
match the leading spin-0 dynamical Love number. 
Finally, we compute 
the running of the scalar
dissipation operators thus extending the previous 
calculations from~\cite{Page:1976df,Goldberger:2020fot,Saketh:2022wap,Saketh:2023bul}.
Our results set the stage for forthcoming spin-2 calculations.

\textit{Worldline EFT and power counting.}--
The first ingredient of the worldline EFT is 
the ``bulk'' action for the massless scalar and gravitational fields 
\be
S_{\rm bulk}=\int d^{4}x\sqrt{-g}\left(\frac{R}{16\pi G} 
- \frac12 (\partial_\mu \phi)^2\right)\,.
\label{eq:bulkaction}
\ee
A compact object of mass $m$ is described by the worldline action
\be
S=-m\int d\tau + S_{\rm fs} \,,
\label{eq:wlaction}
\ee
where $\tau$ is proper time. The first term is the relativistic point-particle action and $S_{\rm fs}$ is an action encoding finite size effects. As mentioned in the introduction, in the language of effective field theory the latter appear as higher-dimension operators on the worldline and couplings of the fields to dynamical multipole moments describing the 
internal degrees of freedom of the compact object, $Q_L$.
For scalars, this action reads~\cite{Goldberger:2005cd,Goldberger:2020fot}
\be 
\label{ac:diss}
S_{\rm fs} =  \sum_
\ell  \int d\tau ~Q_L \bm \d_L \phi 
+ S_{\rm fs}^{\rm ct}  
\,,
\ee 
where in the EFT the multipoles, $Q_L$, are composite operators, $L$ is a multi-index denoting the symmetric traceless combination  $\ell$ indices, and $\bm \d = (g^{\mu\nu}+u^\mu u^\nu) \d_\mu$, with $u^\rho$ the object's 4-velocity, is the spatial derivative in the rest frame of the compact object. $S_{\rm fs}^{\rm ct}$
is the counterterm action 
discussed shortly. The dynamical dipole coupling $\int d\tau {\bm Q}\cdot \bm \d \phi$  is analogous to the familiar dipolar electromagnetic interaction.
In the EFT we are ignorant about the microscopic nature of the multipoles. Instead, we are interested in their correlation functions, such as the 
Fourier transformed time-ordered two-point function,
\be 
\begin{split}
& \int dt e^{-i\omega t}\langle T Q_{L_1}(t)Q_{L_2}(0)\rangle = -i \delta_{L_1L_2} F_\ell(\omega)\,,
\end{split}
\ee
which at low frequencies takes the form
\be
F_\ell(\omega) = \Cs{\ell,\omega^0} + i \Cs{\ell,\omega} |\omega| + \Cs{\ell,\omega^2} \omega^2 + \cdots
\ee
The corresponding Wilson coefficients $\Cs{\ell,\omega^n}$ are 
collectively known as Love numbers. The static Love numbers $(n=0)$ describe the  response of the compact object to time-independent fields (static tides) with different multipolar profile. These have been extensively studied for neutron star and black holes~\cite{Damour:1998jk,Flanagan:2007ix,Damour:2009vw,PoissonWill2014}, which yielded a surprising result that they vanish for black holes in $D=1+3$~\cite{Damour:2009va,Binnington:2009bb,Kol:2011vg,Hui:2020xxx,Chia:2020yla,Charalambous:2021mea} (symmetry explanations were proposed in~\cite{Charalambous:2021kcz,Charalambous:2022rre,Hui:2021vcv,Hui:2022vbh}).
The coefficients $C_{\ell,\omega^{2n}}$  ($n> 0$) are called ``dynamical Love numbers,'' as they describe the response to time-dependent fields. 
We refer to $C_{\ell,\omega^{2n+1}}$ 
as dissipation numbers.

Real parts of $F_\ell$ are analytic functions that
describe conservative finite-size
effects. As such, they can be fully absorbed into 
the local worldline counterterm action,
\begin{align}
 &S_{\rm fs}^{\rm ct}   
 \!=  \!\sum_\ell \frac{1}{2}\!\int  d\tau \Big[ 
  \Cs{\ell,\omega^0}(\bm \d_{\!L}\phi)^2 \!+ \!\Cs{\ell,\omega^2} (\bm \d_{\!L}\dot\phi)^2 \!+ \! \cdots \Big] \\
 & =\frac{1}{2}\int
 d\tau
 \Big[ 
\Cs{1,\omega^0}  (\bm \d \phi)^2 +\Cs{0,\omega^2}\dot\phi^2
+ \Cs{1,\omega^2}(\bm \d \dot \phi)^2
+\cdots \Big]\,, 
\nonumber
\label{eq:fsscalar}
\end{align}
where $\dot \phi= \partial_\tau \phi = (u^\mu \partial_\mu)\phi$. Here in the first line we show operators corresponding to the static and leading dynamical Love numbers, and in the second we show only the leading order 
operators relevant for our calculation below. Note that the scalar monopole operator $\int d\tau \phi^2$ is forbidden by the shift symmetry $\phi \to \phi + v$ of the massless scalar, i.e., $\Cs{0,\omega^0}=0$.

In contrast, $\text{Im}F_\ell$ describes the dissipative part of the 
response and cannot be written in terms of local worldline operators. This will 
capture inelastic effects, e.g. absorption or tidal heating. Note that formally
$F_\ell$ 
are correlators
of renormalized multipole  
moments that receive scale-dependence
through gravitational dressing~\cite{Goldberger:2009qd,Goldberger:2012kf}.

Wilsonian naturalness
dictates that
    \mbox{$\Cs{\ell,\omega^n}\sim R^{2\ell+1+n}$}
where $R$ is the size of the compact object, $R= A G m$, with 
$A=2$ for black holes, and $A\sim 10$
for neutron stars~\cite{Flanagan:2007ix},
which makes it  natural 
to consider $R$ as a second 
expansion parameter independent 
of $Gm$.

\textit{EFT Scattering Amplitudes.}-- 
The effective action~\eqref{eq:bulkaction}-\eqref{eq:wlaction}
can be used to calculate the quantum amplitudes 
for scalars 
scattering off a compact object.  
We compute the full EFT amplitudes following the approach of~\cite{Cheung:2023lnj, JulioNew}, i.e. expanding the effective action around a background solution given by the worldline moving in a straight trajectory, $x^\mu(\tau)=u^\mu \tau$, and the metric given by the large-distance 
expansion of the Schwarzschild metric 
in isotropic coordinates,
\begin{align}
    \bar g_{\alpha\beta} & =  \eta_{\alpha\beta} -   u_\alpha u_\beta \left( -  \frac{\mu}{r^{D-3}} + \frac12 \frac{\mu^2}{r^{2(D-3)}}   \right) \\ 
    & + (\eta_{\alpha\beta} + u_\alpha u_\beta) \left(\frac{1}{D-3} \frac{\mu}{r^{D-3}} - \frac{D-7}{8(D-3)^2} \frac{\mu^2}{r^{2(D-3)}} \right)  \nonumber
    \\ & +\cdots \,,\nonumber
\end{align}
where $\mu=\frac{16 \pi G M}{(D-2) \Omega_{D-2}}$,  and $\Omega_{D-2}$ is the volume of the $(D-2)$-dimensional sphere.  Famously, this expansion resums the perturbative solution to Einstein's equation with a point source~\cite{Duff:1973zz}, which corresponds to an infinite number of worldline Feynman diagrams. The recoil of the worldline is subleading in the low-frequency limit, and the metric fluctuations are not relevant for the scalar field amplitude, so we will also ignore them henceforth.

The full scalar amplitude is simply given by iterative scattering against the background plus the scattering off the dynamical multipole moments. The corresponding background-field Feynman diagrams are
\begin{equation}
\begin{aligned}
\label{eq:totalAmp}
    i{\cal M} & = \vcenter{\hbox{\backgroundone}} + \vcenter{\hbox{\backgroundtwo}}  \\
    & \quad + \vcenter{\hbox{\backgroundQQ}} + \vcenter{\hbox{\backgroundQQoneR}} \\
    & \quad + \vcenter{\hbox{\backgroundQQoneL}} + \cdots\,.
\end{aligned}
\end{equation}
The first background-field vertex Feynman rule is given in momentum space by the Fourier transform of the scalar action
\begin{align}
 \vcenter{\hbox{\backgroundonevertex}} &=   i (\sqrt{-\bar g} \bar g^{\mu\nu}(q) - \eta^{\mu\nu}) k_{1\mu} k_{2\nu}
\end{align}
with momentum transfer $q=k_1+k_2$. The other vertex is just the correlator $\langle Q_{L_1} Q_{L_2} \rangle$ that captures the dynamical multipolar tidal response 
\begin{align}
 &\vcenter{\hbox{\backgroundQQvertex}}= i (-1)^\ell k_{1}^{L_1} k_{2}^{L_2} \delta_{L_1L_2} F_{\ell}(u\cdot k_1)
\end{align}
They are connected by ordinary flat-space propagators 
\begin{align}
\vcenter{\hbox{\scalarprop}}  & = \frac{{-}i}{k^2-i 0}\,.
\end{align}
The background-field diagrams can be recast in terms of ordinary flat-space Feynman integrals~\cite{Cheung:2023lnj, JulioNew}. In our case, at 3PM, all such integrals belong to the family 
\begin{equation}\label{eq:integralfamily}
    G_{a_1a_2a_3a_4a_5a_6a_7} = \int_{\ell_1\ell_2} \frac{\delta(u\cdot \ell_1) \delta(u\cdot \ell_2)D_7^{-a_7}}{D_1^{a_1}D_2^{a_2} D_3^{a_3} D_4^{a_4} D_5^{a_5} D_6^{a_6} }\,,
\end{equation}
with a basis of propagators/invariant products
\begin{equation}
    \begin{aligned}
&D_1 =\ell_1^2, \quad D_2 =\ell_2^2, \quad  D_3=\left(\ell_1+k_1\right)^2, \quad \\
&D_4=\left(\ell_2+k_2\right)^2, \quad
D_5=\left(\ell_1+\ell_2+k_1+k_2\right)^2, \quad  \\ &D_6=\left(\ell_1+k_2\right)^2, \quad D_7=\left(\ell_2+k_1\right)^2  ~.
\end{aligned}
\end{equation}
We will compute all integrals in dimensional regularization with $D=4-2\epsilon$. Using integration-by-parts (IBP) identities \cite{Chetyrkin:1981qh} we can reduce any integral in such family to a basis of \emph{master integrals} given by
\begin{align}
\begin{split}    
\{G_{0011000},  G_{0110100},
G_{1001100},  G_{1100100}, \\
G_{1101100},  G_{1110100},
G_{1111100},  G_{2111100}\}\,.
\label{eq:masterints}
\end{split}
\end{align}
where we use the notation in Eq.~\eqref{eq:integralfamily}.
Their dependence on the frequency $\omega$ is fixed by the dimensional analysis, so they are only non-trivial functions of the scattering angle, which we parameterize by $x=\sin\tfrac\theta 2$. We compute the dependence on $x$ by using the method of differential equations for Feynman integrals \cite{Kotikov:1990kg,Remiddi:1997ny,Gehrmann:1999as}. Indeed, it is not difficult to find a basis $\vec f = \{f_{i=1,\cdots,8}\}$ which satisfies canonical differential equations \cite{Henn:2013pwa,Henn:2014qga}
\begin{equation}\label{eq:diffeq}
    \frac{d\vec f}{dx} = \epsilon \mathbb{A}(x)\vec f = \epsilon \left(\frac{A_0}{x}+ \frac{A_1}{x-1} + \frac{A_{-1}}{x+1}\right)\vec f
\end{equation}
with matrices $A_i$ independent of $x$ and $\epsilon$. The solution easily obtained order by order in $\epsilon$ (see e.g.,~\cite{Henn:2014qga})
\begin{equation}
    \vec f(x, \epsilon) = \vec{f}(0,\epsilon) + \epsilon\int_0^x dx'  \mathbb{A}(x') \vec{f}(0,\epsilon) + \cdots
\end{equation}
The boundary conditions are fixed by requiring the absence of singularities in the backward limit $x=\pm 1$ and expanding around the forward limit $x=0$~\cite{future}.

\textit{Results.}-- Since the worldline operators in~\eqref{eq:fsscalar}
furnish irreducible representations of the rotation group,
it is natural to consider scattering amplitudes in the 
partial wave basis:
\begin{align} 
    i\mathcal{M}(\omega, \theta ) = \frac{2\pi}{\omega}\sum_{\ell =0}^\infty & 
(2\ell+1)(\eta_\ell e^{2i\delta_{\ell}} -1) 
     P_\ell(\cos\theta) \,,\label{eq:pwaves}
\end{align}
where $\theta$ is the scattering angle and $P_\ell$ are Legendre polynomials, 
see Supplemental Material for 
a derivation of our partial 
wave expansion. 
The partial wave coefficients are parameterized in terms of real scattering phase shifts $\delta_{\ell}$ and 
inelasticity parameters $\eta_\ell$ (equivalently, $\Delta\eta_\ell\equiv 1-\eta_\ell$).
In our basis, an operator with a multipole number $\ell$
contributes only to the $\ell$'th wave.

Using $\lambda\equiv 2Gm\omega$,
our final 3PM EFT phase shift for the $\ell$-wave
can be written as
\be 
\label{eq:eftdeltaell}
\begin{split}
&\delta_\ell\Big|_{\rm EFT} =
-\frac{\l}{2\epsilon_{\rm IR}}+\frac{\l}{2}\ln\left(\frac{4\omega^2}{\bar \mu_{\rm IR}^2}\right)
+\sum_{n=1}^3\nu^\ell_n \l^n +\delta^{G^3}_\ell\,,\\
&\Delta\eta_\ell\Big|_{\rm EFT} =
\frac{\ell!\omega^{2\ell+1}\text{Im}F_\ell(\omega)}{2\pi (2\ell+1)!!}
\left(1+ \pi \lambda +\lambda^2\eta^{G^2}_\ell \right)\,,
\end{split}
\ee
where $\nu^\ell_n$  are $\mathcal{O}(1)$ numerical constants, e.g. 
\be 
\nu_2^\ell = 
\frac{-11+ 15\ell(1 + \ell)}{4(-1 + 2\ell)(1+2\ell)(3+2\ell)} \pi\,,
\ee
and the 
rest are given in Supplemental Material.
$\bar\mu_{\rm IR}$
is the IR matching scale. 
$\delta_\ell^{G^3}$ and $\eta_\ell^{G^2}$
contain UV divergences.
Elastic terms $\delta_\ell^{G^3}$ are non-zero only for $\ell=0,1$:
\begin{align}
\nonumber
   &  \delta_0^{G^3}\Big|_{\rm EFT} = \l^3
\left[ \frac{1}{4 \epsilon_{\rm UV}} + \frac{13}{6}  - \frac{1}{2} \ln\left( \frac{4\omega^2}{\bar \mu^2}\right)\right]  +\frac{\Cs{0,\omega^2}\omega^3}{4\pi} \,,\\
\label{eq:Swave}
 &   \delta_1^{G^3}\Big|_{\rm EFT} =  
    \frac{\Cs{1,\omega^0}\omega^3}{12\pi}\left(1+ \pi\lambda +\lambda^2\eta^{G^2}_1\right) + \frac{\Cs{1,\omega^2}\omega^5}{12\pi} \,,
\end{align}
where $\bar\mu$
is the matching scale
in the minimal subtraction ($\overline{{\rm MS}}$) scheme. 
The expression for $\eta^{G^3}_\ell$
is given in the Supplemental Material. 
For $\ell=0,1$ we have:
\be 
\begin{split}
& \eta^{G^2}_0\Big|_{\rm EFT} = 
 \frac{67}{12}-\frac{11}{6}\left(-\frac{1}{2\epsilon_{\rm UV}} + \ln\left(\frac{4\omega^2}{\bar\mu^2}\right) \right)+\frac{\pi^2}{3}\\
& \eta^{G^2}_1 \Big|_{\rm EFT}=  \frac{413}{100}
-\frac{19}{30}\left(-\frac{1}{2\epsilon_{\rm UV}} + \ln\left(\frac{4\omega^2}{\bar\mu^2}\right) \right)+\frac{\pi^2}{3}\,.\\
\end{split}
\ee 
The IR divergence and the $\bar\mu_{\rm IR}$ dependence in the first two terms of Eq.~\eqref{eq:eftdeltaell} are unobservable~\cite{Weinberg:1965nx,Amati:1990xe} because they appear multiplicatively 
in the S matrix and hence do no affect the physical cross-section.   
The third term is a sum of the finite 
Feynman diagrams. The last term in the first line of Eq.~\eqref{eq:eftdeltaell}
contains the UV singularity in the single insertion of the background metric at $O(G^3)$
and the relevant tree-level worldline counterterms (Love numbers), displayed in Eq.~\eqref{eq:Swave}.
Since there are no divergences in the P-wave at $\mathcal{O}(G^3)$,
$\Cs{1,\omega^0}$ is just a constant. 
In contrast, 
$\Cs{0,\omega^2}(\bar\mu)$ is a running coupling, which we use
to renormalize the S-wave divergence in the $\overline{{\rm MS}}$
scheme. Its $\beta$-function is given by 
\be
\frac{d 
\Cs{0,\omega^2}
(\bar\mu)}{d\ln \bar\mu} =-4\pi (2Gm)^3\,.
\ee
It is convenient now to absorb all local counterterms 
into the real part of internal multipole
moments defined in Eq.~\eqref{ac:diss}. 
Then we can write down a unified 
expression for the two-loop beta function of all scalar 
tidal operators: 
\be 
\frac{d F_\ell(\omega;\bar\mu)}{d\ln\bar\mu} =- (2Gm\omega)^2\left[\frac{4\nu_2^\ell}{\pi}  F_\ell(\omega;\bar\mu)  +8\pi Gm \delta_{[0\ell]}\right]
\,,
\ee 
where $\delta_{[\ell\ell']}$ is
the Kronecker delta. 
The first term in the r.h.s. above describes the 
running of self-induced tidal effects, both conservative 
and dissipative. The EFT elegantly explains
this homogeneity: both effects stem from the correlators
$\langle QQ\rangle$ that pick up the same running 
 from gravitational 2-loop diagrams attached to them.
Interestingly, the two-loop beta function is proportional to the one-loop (2PM) phase shift $\nu_2^\ell$. This can be explained by the fact that unitarity fixes the coefficient of the $\ln \omega^2$ in the dressed correlator in terms of the lower-PM amplitude. 
In contrast, the rightmost term above is a \textit{universal} 
conservative contribution which arises 
from the PM expansion. The EFT dictates that 
this part does not depend on the nature of the compact object.

\textit{Matching to black holes.}-- Let us compare our EFT phase shift~\eqref{eq:eftdeltaell} 
with the analytic expression known from BHPT in GR~\cite{1978ApJS...36..451M,1988sfbh.book.....F,Mano:1996vt,Mano:1996mf,Mano:1996gn,Sasaki:2003xr,Dolan:2008kf,Bautista:2023sdf}.
Truncating this 
expression at 3PM and 
introducing the Schwarzschild radius $r_s=2Gm$
as the only scale of static black holes,
we find
\be
\label{eq:bhpt_spin0}
\delta_\ell\Big|_{\rm GR} = 
 (r_s\omega)\ln\left(2{\omega}r_s\right)
+  \sum_{n=1}^3 {\nu}^{\ell}_n (r_s \omega)^n +\delta_\ell^{G^3}
\,,
\ee
where $\delta_\ell^{G^3}\Big|_{\rm GR}=0$ for $\ell>0$ and
\be 
\label{eq:bhptdinLNlog}
\delta_0^{G^3}\Big|_{\rm GR}=(r_s\omega)^3 \left[ \frac{7}{12} -\g_E - \ln(2r_s\omega) \right]\,,
\ee 
while the 3PM 
inelasticity parameters
are given by 
\be 
\Delta\eta_\ell\Big|_{\rm GR} =
\small{\frac{2^{2\ell+1}(\ell!)^4(r_s \omega)^{2\ell+2}}{[(2\ell)!(2\ell+1)!]^2(2\ell+1)}}
\left(1+ \pi \lambda +\lambda^2\eta^{G^2}_\ell
\right)\,,
\ee
with
\be 
\begin{split}
\label{eq:bhptlog}
&\eta_0^{G^2}\Big|_{\rm GR}=-
    \frac{11}{6}\ln\Bigg(4 r_s^2 \omega^2 e^{2\gamma_E} \Bigg)+ \frac{{2}\pi^2}{3} + \frac{191}{36}\,, \\
   &\eta_1^{G^2}\Big|_{\rm GR}=-
   \frac{19}{30}
    \ln\Bigg(4 r_s^2 \omega^2 e^{2\gamma_E} \Bigg)+ \frac{{2}\pi^2}{3} + \frac{6853}{900} \,,
\end{split}
\ee 
and the rest are given in Supplemental Material.


As a first check of our calculation, we verify that
infrared divergences in the EFT match those in the full theory by choosing $\bar \mu_{\rm IR} = 1/r_s$.
A second important 
observation is that the coefficients in front of the UV logs in the EFT
expression~\eqref{eq:Swave} match those in GR (\ref{eq:bhptdinLNlog},\ref{eq:bhptlog}), as expected by consistency of the EFT.
Matching the P-wave ($\ell = 1$) phase shift we obtain the vanishing of the scalar dipole static Love number, $C_{1,\omega^0}=0$,
consistent with previous results~\cite{Kol:2011vg,Hui:2020xxx,Charalambous:2021mea,Ivanov:2022qqt}. This is 
the first rigorous on-shell proof
of the vanishing of Love numbers.
The contribution of the dipolar dynamical Love number
$C_{1,\omega^2}$ in Eq.~\eqref{eq:Swave}
shifts to 5PM 
for black holes,
which is beyond the scope
of our work. 

Matching the S-wave, we extract the
monopole dynamical Love number:
\be
\Cs{0,\omega^2}
(\bar\mu)^{\overline{{\rm MS}}} = -4\pi r_s^3 \left[\frac{1}{4\epsilon_{\rm UV}}+\ln(\bar\mu r_s)+\frac{19}{12}+\g_E\right]\,,
\ee
obtained in the conventional dimensional regularization + $\overline{{\rm MS}}$ scheme.
This is one of our main results. Note it is broadly consistent with  the numerical estimate from~\cite{Barack:2023oqp}.

Finally, matching $\eta_\ell$ we get
the 
renormalized
$\text{Im}F_\ell$, e.g. for the S-wave
we have 
\be 
\begin{aligned}
\text{Im}F_0(\omega;\bar\mu)^{\overline{{\rm MS}}} &  = 
4\pi r_s^{2} |\omega|\bigg(1 + (r_s \omega)^2
\bigg[\frac{\pi^2}{3}-\frac{5}{18}
 \\
& -\frac{11}{3}\bigg(
\ln\left(\bar\mu r_s\right) +\g_E \bigg) \bigg]\bigg)~\,.
\end{aligned}
\ee 

Concluding, we note that comparison with GR demonstrates the utility 
of the EFT. 
Although the full GR phase shift 
is known, 
the physical interpretation 
of individual terms in it,
especially logarithms, is difficult. 
In contrast, the EFT clearly classifies all logs into IR and UV ones, and also distinguishes 
universal and self-induced tidal effects. Finally, 
the EFT nicely explains  
the apparent conspiracy between
coefficients 
in front of dissipative logs and
conservative phase shifts at lower loop orders.

\textit{Generalizations.--} 
Our method can be used to study
tides in higher 
 spacetime dimension $D$. 
For general $D$, it is trivial to match 
finite-size couplings because tidal effects
do not scale as integer powers of $G$~\cite{Kol:2011vg,Hui:2020xxx, Ivanov:2022qqt,Saketh:2023bul}.
UV divergences and non-trivial matching
conditions arise if $2\ell/(D-3)$ is integer.
In particular, in $D=5$ we find divergences for both S and P-waves,
\be 
\delta_\ell\Big|_{\text{EFT}}^{D=5} \supset -\frac{(Gm\omega^2)^2}{72\pi}\left( 64\delta_{[\ell 0]}  + \delta_{[\ell 1]}\right)\ln\left(\frac{\omega}{\bar\mu}\right)
\,.
\ee  
Their renormalization requires 
the following universal running of the
worldline couplings 
\be 
\frac{d \Cs{1,\omega^0}}{d\ln \bar\mu}=-\frac{8}{9}(Gm)^2\,,\quad 
\frac{d \Cs{0,\omega^2}}{d\ln \bar\mu}=-\frac{128}{9}(Gm)^2\,.
\ee 
The beta-function for $C_{1,\omega^0}$
matches the known results from GR~\cite{Kol:2011vg,Hui:2020xxx,Ivanov:2022qqt}. 
The running part of the dynamical LN 
$C_{0,\omega^2}$ is obtained
for the first time.
Since full BHPT results are 
not readily available in the literature
for \mbox{$D=5$},
we leave a complete matching for future 
work.

\textit{Conclusions.--}We have presented a new systematic framework
to match tidal responses 
of compact objects 
from probability amplitudes 
of massless waves to scatter 
off these objects. 
Our method 
is free of gauge dependence 
and field-redefinition ambiguities 
that plague the standard
off-shell matching techniques commonly used to 
extract tidal effects (Love numbers). 
We illustrated the power 
of our approach 
by calculating 
a full 3PM 
amplitude 
for a scalar field to 
quasi-elastically 
scatter 
off a generic compact object.
Our technique 
leads to rich implications for black holes, 
for which analytic GR 
results
are available for comparison.
In particular, 
we clarified the IR and UV origin 
of different terms 
in the GR expressions.
Overall, our findings 
presented here give new 
insights
into the form of gravitational  scattering
amplitudes,
and serve as a prototype  
for the upcoming spin-2 
Raman scattering 
calculations. 

\emph{Note added}: The previous version of this paper contained an error in the constant part of the S-wave 3PM dynamical Love number.  The error originated from a subtlety in evaluating the S-wave phase shift. In particular, formulas for general $\ell$  cannot be trusted at integer $\ell < \ell^\ast$ if $\ell^\ast$ appears as a Regge pole, even if there is no singularity in the partial wave at such integer $\ell$. In our case, there is a Regge pole at $\ell^\ast = 1/2$, invalidating the naive analytic continuation of the general-$\ell$ formula to $\ell = 0$. This gave rise to a discrepancy between the (mistakenly) analytically-continued general-$\ell$ result to $\ell = 0$ and the direct calculation of the S-wave phase shift. We are grateful to M. Correia, S. Caron-Huot, G. Isabella, and M. Solon for pointing out a disagreement with their upcoming work, which lead us to identify this error.

\textit{Acknowledgments.}
We thank Giulio Bonelli, Simon Caron-Huot, Clifford Cheung, Horng Sheng Chia,  Alfredo Guevara, Hofie Hannesdottir, Cristoforo Iossa, Henrik Johansson, Gregor K{\"a}lin, Alex Ochirov, Rafael Porto, Muddu Saketh, Nabha Shah, Alessandro Tanzini and Jordan Wilson-Gerow for insightful discussions; and specially Yilber Fabian Bautista, Thibault Damour, Jung-Wook Kim and Ira Rothstein for discussions and comments on the draft. YZL is supported by the US National Science Foundation under Grant No. PHY- 2209997.

\bibliography{short.bib}

\newpage 

\pagebreak
\widetext
\begin{center}
\textbf{\large Supplemental Material}
\end{center}
\setcounter{equation}{0}
\setcounter{figure}{0}
\setcounter{table}{0}
\setcounter{page}{1}
\makeatletter
\renewcommand{\theequation}{S\arabic{equation}}
\renewcommand{\thefigure}{S\arabic{figure}}
\renewcommand{\bibnumfmt}[1]{[S#1]}
\renewcommand{\citenumfont}[1]{S#1}

\section{Partial wave expansion}

In this appendix, we derive the partial wave expansion for the quantum amplitude of scalar fields scattering off a compact object. The four-dimensional partial waves contain infrared divergences which we regularize by working in $D=4-2\epsilon_{\rm IR}$ dimensions, i.e., we will work with the analog of \eqref{eq:pwaves} in general dimension, $D$.

We consider the following scalar $1$-to-$1$ S-matrix on the background sourced by the compact object
\be
\langle k_2|(S-1)|k_1 \rangle = i \mathcal{M}(k_1 \rightarrow k_2) \cdot (2\pi) \delta\left(u \cdot (k_1 + k_2)\right) ~.
\ee
The direction of $k_i$ transverse to $u$ is denoted by $n_i$. By projecting the above amplitude onto the traceless symmetric irreducible representation $\rho$ of SO$(D-1)$ using the projector $| L \rangle \langle L |$, we obtain the partial wave expansion
\be
i \mathcal{M}(k_1 \rightarrow k_2)=\sum_{\ell} (\eta_\ell e^{2i\delta_\ell}-1) \langle n_2 |L \rangle \langle L |n_1 \rangle \, ~.
\ee
The vertex $\langle L |n \rangle$ can be fixed by the ${\rm SO}(D-1)$ symmetry up to a kinematic normalization $\mathcal{N}(\omega)$
\be
\langle L|n\rangle=\Bigg[\mathcal{N}(\omega)\frac{2^\ell \big(\frac{D-3}{2}\big)_\ell}{(D-3)_\ell}\Bigg]^{\frac{1}{2}} \Big(n_{\mu_1}\cdots n_{\mu_\ell}-\text{traces}\Big)\,.
\ee
Here, $(a)_n$ is the pochhammer symbol.
Explicit calculation shows that
\be
\langle n_2|L \rangle \langle L |n_1\rangle= \mathcal{N}(\omega) P_\ell^{(D)}(\cos\theta)\,,
\ee
where $P_\ell^{(D)}(z)$ is the Gegenbauer function (polynomials for $\ell \in \mathbb{N}$)
\be
P_\ell^{(D)}(z)=\,_2F_1\Big(-\ell,\ell+D-3,\frac{D-2}{2},\frac{1-z}{2}\Big)\, ,
\ee
and
${n_1 \cdot n_2} = z= \cos\theta$. 

To determine the normalization, we recall that the normalization of one-particle state is $\langle k_2|k_1\rangle=2\omega (2\pi)^{D-1}\delta^{D-1}(k_1-k_2)$, thereby yielding the following completeness relation
\be
\frac{(2\omega)^{D-3}}{(4\pi)^{D-2}}\int_{S^{D-2}}dn|n\rangle \langle n|\equiv 1\,.
\ee
Given that the dimension of the irreducible representation can be determined by the group trace of the projector ${\rm Tr}\,|L\rangle \langle L|$, we obtain the following relation
\be
{\rm dim} \rho = \frac{(2\ell+D-3)\Gamma(D+\ell-3)}{\Gamma(D-2)\Gamma(\ell+1)}={\rm Tr}\,|L\rangle \langle L|=\frac{(2\omega)^{D-3}{\rm Vol}\,S^{D-2}}{(4\pi)^{D-2}} \mathcal{N}(\omega)\,.
\ee
The above relation then completely determines the normalization in the partial wave expansion, and we end up with
\begin{align} 
    i\mathcal{M}(\omega, \theta ) = \sum_{\ell =0}^\infty 
\frac{n_{\ell}^{(D)}}{(2\omega)^{D-3}} (\eta_\ell e^{2i\delta_{\ell}} -1) 
     P_\ell^{(D)}(\cos\theta) \,,\quad n_{\ell}^{(D)} = \frac{(4\pi)^{\frac{D-2}{2}}(D+2\ell-3)\Gamma(D+\ell-3)}{ \Gamma(\frac{D-2}{2})\Gamma(\ell+1)}\,.
\end{align}
Using the orthogonality relation
\be
 \int_{-1}^{+1} dz \left(1-z^2\right)^{\frac{D-4}{2}} P_\ell^{(D)} (z) P_{\ell'}^{(D)} (z)  = \frac{\Gamma(\frac{D-2}{2})}{2\left(16 \pi\right)^{\frac{2-D}{2}}} \frac{\delta_{[\ell\ell']}}{n_{\ell}^{(D)}}\,,
\ee
we find that the partial wave coefficients are given by
\begin{align}\label{eq:partl}
 (\eta_\ell e^{2i\delta_{\ell}} -1) = \frac{i\omega}{2\pi} \frac{\left(4 \pi/\omega^2\right)^{\frac{4-D}{2}}}{ 2\Gamma(\frac{D-2}{2})} \int_{-1}^1 d z\left(1-z^2\right)^{\frac{D-4}{2}} P_\ell^{(D)} (z) \mathcal{M}(\omega, z)\,.
\end{align}

As already mentioned, the amplitude contains both infrared divergences and singularities in the forward limit, which are regularized by working in $D=4-2\epsilon_{\rm IR}$ dimensions \footnote{Note that the UV divergence is regulated by $D= 4 -2 \epsilon_{\rm UV}$ and we do not keep terms with $\mathcal{O}(\epsilon_{\rm IR}/\epsilon_{\rm UV})$}.
The partial wave transform of power-law forward singularities can be obtained for general $\ell$ via the inversion formula 
\begin{align}
 \quad \frac{\left(4 \pi \mu_{\rm IR}^2\right)^{\epsilon_{\rm IR}}}{2 \Gamma(1-\epsilon_{\rm IR})} \int_{-1}^1 d z\left(1-z^2\right)^{-\epsilon_{\rm IR}} P_\ell^{(D)} (z) \left(\frac{1-z}{2}\right)^p \nonumber  =  (\pi\mu_{\rm IR}^2)^{\epsilon_{\rm IR}} \frac{\Gamma(1-\epsilon_{\rm IR} + p)\Gamma(\ell-p)}{\Gamma(-p)\Gamma(2-2\epsilon_{\rm IR} + \ell + p)}\,.
\end{align}

Some care is required when performing the partial wave expansion of logarithmically divergent terms, e.g. $\ln(z-1)$.
More concretely, the limit $\epsilon_{\rm IR} \to 0$ does not commute with the forward limit $z\to 1$. Fortunately, the differential equations in Eq.~\eqref{eq:diffeq}, provide a systematic way of expanding around the forward limit with fixed $\epsilon_{\rm IR}$, with the leading behavior $\vec f \sim x^{\epsilon_{\rm IR} A_0} \sim (z-1)^{\epsilon_{\rm IR} A_0}$.

The above expression 
is to be used in conjunction with the 
analytical continuation of $\ell$ \cite{Kol:2011vg,Charalambous:2021kcz,Creci:2021rkz,Ivanov:2022qqt,Bautista:2023sdf} valid for sufficiently large 
orbital numbers 
thanks to the 
unitarity.
This can be seen from the Froissart-Gribov formula~\cite{gribov2002gauge}, which not only proves the analyticity for $\ell \geq \ell_0$, but also provides a shortcut to compute the phase shift from the $t$-channel cut (where $t=2\omega^2(z-1)$) of the amplitudes
\begin{align}\label{eq:partl}
 (\eta_\ell e^{2i\delta_{\ell}} -1) = \frac{i\omega}{\pi^2} \frac{\left(4 \pi/\omega^2\right)^{\frac{4-D}{2}}}{ 2\Gamma(\frac{D-2}{2})} \int_{1}^\infty d z\left(z^2-1\right)^{\frac{D-4}{2}}Q_\ell^{(D)} (z) {\rm Disc}_t \mathcal{M}(\omega, z)\,, 
\end{align}
where $Q_\ell^{(D}(z)$ is the Gegenbauer-Q function
\be
\label{eq:FG}
Q_{\ell}^{(D)}(z)=\frac{\sqrt{\pi}\Gamma(\ell+1)\Gamma(\frac{D-2}{2})}{2^{\ell+1}\Gamma(\ell+\frac{D-1}{2})}\frac{1}{z^{\ell+D-3}} \,_2F_1\Big(\frac{\ell+D-3}{2},\frac{\ell+D-2}{2},\ell+\frac{D-1}{2},\frac{1}{z^2}\Big)\,.
\ee
The Froissart-Gribov representation is well suited for an expansion around large $\ell$, which is dominated by the lower limit of the integral, i.e., $z\to 1$, and hence conveniently computed given the forward expansion of the amplitude. Furthermore, as it only features the discontinuity in $z$ of the amplitude, the computation of the general-$\ell$ partial waves is greatly simplified. For instance, out of the eight master integrals in Eq.~\eqref{eq:masterints} only the four in the second line are non-analytic in the scattering angle, and hence contribute to the discontinuity. The restriction $\ell \geq \ell_0$ arises from the underlying Regge limit of the amplitudes in the complex $t$-plane, characterized by $\lim_{|t|\rightarrow\infty}|\mathcal{M}|<|t|^{\ell_0}$. This restriction appears during the contour deformation of the amplitudes in the complex $t$-plane when deriving the Froissart-Gribov formula, because it is necessary for the integrand at complex infinity to decay sufficiently fast so that it can be dropped given $Q_\ell^{(D)}\Big|_{z\rightarrow\infty}\sim 1/z^{\ell+D-3}$ (see e.g. \cite{Correia:2020xtr}). Let us note that the Froissart-Gribov formula produces spurious poles 
for physical $\ell \in \mathbb{N}$ at $\mathcal{O}(G^{2\ell+3})$. These are similar to the spurious poles encountered in GR calculations of static Love numbers
or higher dimensional Schwarzschild 
black holes in 
\cite{Kol:2011vg}
and dynamical Love numbers of Kerr black holes in \cite{Charalambous:2021mea}.
In that case, we use the
partial wave transformation for fixed integer $\ell$.

\section{Partial wave amplitudes for general $\ell$-waves}

In this section, we present expressions for general $\ell$'th partial waves. Note that we use the optical theorem to relate the inelasticity coefficients and ${\rm Im} F_\ell$ through the absorption cross-section:
\be 
\sigma_{{\rm abs},\ell}=
 \frac{\pi(2\ell+1)}{\omega^2}\left(1-\eta^2_{\ell}\right)= \frac{\ell ! \omega^{2\ell-1} {\rm Im} F_\ell(\omega)}{(2\ell-1)!!} ( 1+ {\cal O}(G)) \,.
\ee 

\begin{table*}[hbt!]
    \centering
    \setlength\tabcolsep{14pt}
    {\tabulinesep=1.0mm
    \begin{tabular}{|c|c|c|c|}
        \hline
         $\nu_n^\ell$ &  $n=1$   & $n=2$  & $n=3$ \\ [0pt]
         \hline
         $\ell=0$ & $- \frac{1}{2} + \gamma_E$  & $\frac{11}{12} \pi$  & $ \frac{11}{36}\pi^2 - \frac{1}{3} \zeta(3) - \frac{1}{12}$ \\ [0pt]
         \hline
         $\ell=1$ & $ - \frac{3}{2} + \gamma_E$ & $\frac{19}{60}\pi$ & $\frac{19}{180}\pi^2 - \frac{\zeta(3)}{3}$  \\ [0pt]
         \hline
         \multirow{2}[0]{*}{generic $\ell$} & \multirow{2}[0]{*}{$- \frac{1}{2} - \psi^{(0)}(1+\ell)$} & 
         \multirow{2}[0]{*}{$\frac{-11+ 15\ell(1 + \ell)}{4(-1 + 2\ell)(1+2\ell)(3+2\ell)} \pi$} & 
         $\frac{1}{2} \frac{-11 + 15 \ell + 15 \ell^2}{(-1 + 2\ell)(1+2\ell)(3+ 2\ell)} \psi^{(1)}(1+\ell) + \frac{1}{6} \psi^{(2)}(1+\ell)$
         \\
        & & &  $+ \Big( \frac{1}{2\ell} - \frac{1}{2(1+\ell)} + \frac{1}{16(-1+2\ell)} - \frac{1}{16(3+2\ell)}\Big)$ \\ [0pt]
         \hline
     \end{tabular}}
    \caption{Coefficients $\nu_n^\ell$ for $\ell=0,1$ and generic-$\ell$ in Eq.~\eqref{eq:Swave}. $\gamma_E=0.577...$ is the Euler-Mascheroni constant, $\zeta(x)$ is the Riemann zeta function, and $\psi^{(i)}(x)$ is the $i$-th order polygamma function.}
    \label{tab:coeff}
\end{table*}

The EFT expression for the 
3PM amplitude for a general $\ell$
can be written as 
\be
\begin{aligned}
\delta_\ell\Big|_{\rm EFT} = 
& -\frac{Gm\omega}{\epsilon_{\rm IR}}+(2Gm\omega) \frac{1}{2}\ln\left(\frac{4\omega^2}{\bar \mu_{\rm IR}^2}\right) +\sum_{n=1}^3\nu^\ell_n (2Gm \omega)^n \\
& +
\delta_{[0\ell]}(2Gm\omega)^3
\left[ \frac{1}{4 \epsilon_{\rm UV}} + \frac{13}{6}  - \frac{1}{2} \ln\left( \frac{4\omega^2}{\bar \mu^2}\right)\right] +
\text{Re}F_\ell(\omega;\bar\mu) \mathcal{F}_{\rm EFT}(\omega)
\,,\\
\eta_\ell\Big|_{\rm EFT}  =& 1 - 
2 \text{Im}F_\ell(\omega;\bar\mu)\mathcal{F}_{\rm EFT}(\omega)
\,,
\end{aligned}
\ee
where the EFT tidal form factor with  attached two-loop gravitational corrections is given by
\be 
\mathcal{F}_{\rm EFT}(\omega) = \frac{\ell!\omega^{2\ell+1}}{4\pi (2\ell+1)!!}\Bigg\{ 
 1+ \pi (2Gm\omega)
+(2Gm\omega)^2
\Bigg[ \frac{4\nu_2^\ell}{\pi}\left(\frac{1}{4\epsilon_{\rm UV}} - 
\frac{1}{2}\ln\left(\frac{4\omega^2}{\bar\mu^2}\right) \right)+\frac{\pi^2}{3}+ d^{\rm EFT}_\ell \Bigg]\Bigg\}\,,
\ee 
the numerical coefficients $\nu^\ell_n$ are given in Table \ref{tab:coeff},
while 
the coefficients $d_\ell^{\rm EFT}$ 
are to be retrieved for 
each integer $\ell$, e.g. $d_0^{\rm EFT} = 67/12$, $d_1^{\rm EFT} = 413/100$, etc.
Note that we have ignored phase contributions to the above inelasticity 
parameters that eventually get canceled in physically measurable
cross-sections. Our convention is $\bar\mu_{\rm IR}^2 = \mu_{\rm IR}^2 4\pi e^{\gamma_E-1}$ and $\bar\mu^2 = \mu^2 4 \pi e^{-\gamma_E}$.
For black holes, our EFT expressions
can be compared with the  
BHPT results from the literature:
\begin{align}
\begin{split}
\delta_\ell\Big|_{\rm GR} = &
 (r_s\omega)\ln\left(2{\omega}r_s\right)
+  \sum_{n=1}^3 {\nu}^{\ell}_n (r_s \omega)^n 
 + \delta_{[0\ell]} (r_s\omega)^3 \Bigg[ \frac{7}{12} - \gamma_E - \ln(2r_s\omega) \Bigg],\\
\eta_\ell\Big|_{\rm GR} = & 1 -
\frac{2^{2\ell+1}(\ell!)^4(r_s \omega)^{2\ell+2}}{[(2\ell)!(2\ell+1)!]^2(2\ell+1)}
    \Bigg\{ 1
    + \pi (r_s \omega) 
    + (r_s \omega)^2 
    \Bigg[ -
    \frac{2\nu^\ell_2}{\pi}\ln\Bigg(4 r_s^2 \omega^2 e^{2\gamma_E} \Bigg)+ \frac{ 2 \pi^2}{3} + d^{\rm BHPT}_\ell \Bigg]\Bigg\}\,,
\end{split}
\end{align}
where $d_0^{\rm BHPT}=191/36$
for the S-wave, while  
for generic $\ell>0$ we have:
\be 
\begin{aligned}
    d_\ell^{\rm BHPT}  = 
    \frac{4\nu_2^\ell}{\pi}\Bigg[ \frac{2}{1+2\ell} - 3 (\psi ^{(0)}(\ell +1)+\gamma_E ) + 4 (\psi ^{(0)}(2 \ell +1)+\gamma_E ) \Bigg]  +\frac{-2 \ell -1}{4 (2 \ell -1)^2 (2
   \ell +3)^2}-3 \psi ^{(1)}(\ell +1)+\frac{\pi ^2}{2} \,,
\end{aligned}
\ee
e.g. $ d_1^{\rm BHPT} = 6853/900, d_2^{\rm BHPT} = 8492/1225$ for P- and D-waves.
Performing a wavefunction renormalization 
of multipoles
in action~\eqref{eq:wlaction} via
a formal replacement 
\be 
Q_{L} \to Q^{\rm bare}_{L} = Q^{\rm ren.}_{L}(\omega;\bar\mu)
Z_\ell(\omega;\bar\mu)\,,
\ee 
with $Z_\ell$ being the 
wavefunction 
renormalization constant, 
the matching of GR 
and EFT results yields
\be 
\begin{aligned}
&\text{Im}F_\ell(\omega;\bar\mu)^{\overline{{\rm MS}}}   = 
\small{\frac{2^{2\ell+2}\pi(\ell!)^3(2\ell-1)!!}{[(2\ell)!(2\ell+1)!]^2}}
|\omega|\bigg(1 +(r_s \omega)^2
\bigg[\frac{\pi^2}{3}+d_\ell^{\rm BHPT}-d_\ell^{\rm EFT}
-\frac{4\nu_2^\ell}{\pi}\bigg(
\ln\left(\bar\mu r_s\right) +\g_E \bigg) \bigg]\bigg)~\,,\\
&Z_\ell(\omega;\bar\mu)^{\overline{{\rm MS}}}  = 1 -(r_s \omega)^2
\frac{\nu_2^\ell}{2\pi}
\frac{1}{\epsilon_{\rm UV}}~\,,
\end{aligned}
\ee 
where $F_\ell(\omega;\bar\mu)^{\overline{{\rm MS}}}$ is the two-point
function of the renormalized 
multipoles $Q_L^{\rm ren.}$.


\end{document}